\documentclass[sigconf]{acmart}

\usepackage[prependcaption,textsize=tiny]{todonotes}

\usepackage{hyperref}
\usepackage{url}

\usepackage{multirow}
\usepackage{todonotes}
\usepackage{float}

\usepackage{graphicx}
\usepackage{caption}
\usepackage{subcaption}

\ccsdesc[500]{Software and its engineering~Software design tradeoffs}
\ccsdesc[500]{Computing methodologies~Machine learning}
\keywords{Machine learning, software engineering, flow-based programming, service-oriented architecture}

\title[]{An Empirical Evaluation of Flow Based Programming in the Machine Learning Deployment Context}

\author{Andrei Paleyes}
\affiliation{%
	\institution{University of Cambridge}
	\department{Department of Computer Science and Technology}
	\country{United Kingdom}
}
\email{ap2169@cam.ac.uk}

\author{Christian Cabrera}
\affiliation{%
	\institution{University of Cambridge}
	\department{Department of Computer Science and Technology}
	\country{United Kingdom}
}
\email{chc79@cam.ac.uk}

\author{Neil D. Lawrence}
\affiliation{%
	\institution{University of Cambridge}
	\department{Department of Computer Science and Technology}
	\country{United Kingdom}
}
\email{ndl21@cam.ac.uk}

\copyrightyear{2021} 
\acmYear{2021} 
\setcopyright{rightsretained} 
\acmConference[CAIN'22]{1st Conference on AI Engineering - Software Engineering for AI}{May 16--24, 2021}{Pittsburgh, PA, USA}
\acmBooktitle{1st Conference on AI Engineering - Software Engineering for AI (CAIN'22), May 16--24, 2021, Pittsburgh, PA, USA}
\acmDOI{10.1145/3522664.3528601}
\acmISBN{978-1-4503-9275-4/21/05}

\newcommand{\change}[1]{#1}

\begin{document}

\begin{abstract}
As use of data driven technologies spreads, software engineers are more often faced with the task of solving a business problem  using data-driven methods such as machine learning (ML) algorithms. Deployment of ML within large software systems brings new challenges that are not addressed by standard engineering practices and as a result businesses observe high rate of ML deployment project failures. Data Oriented Architecture (DOA) is an emerging approach that can support data scientists and software developers when addressing such challenges. However, there is a lack of clarity about how DOA systems should be implemented in practice. This paper proposes to consider Flow-Based Programming (FBP) as a paradigm for creating DOA applications. We empirically evaluate FBP in the context of ML deployment on four applications that represent typical data science projects. We use Service Oriented Architecture (SOA) as a baseline for comparison. Evaluation is done with respect to different application domains, ML deployment stages, and code quality metrics. Results reveal that FBP is a suitable paradigm for data collection and data science tasks, and is able to simplify data collection and discovery when compared with SOA. We discuss the advantages of FBP as well as the gaps that need to be addressed to increase FBP adoption as a standard design paradigm for DOA.
\end{abstract}

\maketitle

\section{Introduction}
\label{section:intro}
When deploying machine learning (ML) algorithms in real-world systems, software developers face a new set of challenges~\cite{paleyes2020challenges, Figalist2022BreakingTV}. In particular, real-world systems produce large quantities of heterogeneous, time varying, high dimensional data that feeds decision making in these systems. This challenges the effectiveness and efficiency of current software development and deployment practices. The challenges are present across the entire ML application workflow, including the stages of data engineering, model learning, model verification, and model deployment. For example, data analysts spend most of their time in looking for, acquiring, understanding, cleaning and preparing the data before using a ML algorithm~\cite{nazabal2020data}. These challenges arise because the ML solutions are deployed on top of existing software solutions which were built to fulfill goals that are important but not directly related to ML, for example high availability, robustness, and low latency. Machine learning poses a new set of challenges that the majority of existing software architectures were not designed for \cite{paleyes2020challenges, lewis2021software, ozkaya2020what}.

This paper considers software architectures that might be more appropriate for converting data into business value. One paradigm proposed in the research community is known as Data Oriented Architectures (DOA). DOA is an emerging software architectural pattern that aims to facilitate the integration of machine learning algorithms within modern software systems \cite{joshi2007data, vorhemus2017data, lawrence2019doa}. DOA treats data in the system as a first class citizen in a shared information model, where stateless system components perform distributed processing. These components communicate between each other using an asynchronous message exchange protocol. Such features enable DOA to achieve high data discoverability, availability, and reuse. The stateless and loosely coupled system components also allow DOA to deal with large-scale dynamic environments~\cite{vorhemus2017data}.

Despite a high level consensus about the potential benefits that DOA brings to the implementation and deployment of ML algorithms, there is no clarity on which tools, frameworks and programming paradigms should be used as the building blocks of a DOA system in practice. This paper presents a quantitative evaluation of flow-based programming (FBP) \cite{morrison1994flow} as a paradigm to use for DOA-based applications. FBP defines applications as  networks of "black box" processes, which communicate via data streaming connections, where the connections are specified externally to the processes. Through these principles, such as external connections and named ports, FBP promotes data coupling between system components. We evaluate to what extent FBP is suitable for the development of ML-based applications at different stages of the ML deployment workflow. As a baseline for comparison in our evaluation we use the currently prevalent Service Oriented Architecture (SOA) \cite{perrey2003service, oreilly2020microservices}.

Our evaluation follows the Goal-Question-Metric (GQM) methodology for experimentation in software engineering proposed by Wohlin et. al.~\cite{wohlin2012experimentation}. We first use the GQM framework to define goals of the evaluation, their respective questions, and the evaluation metrics. We next describe and develop four data processing applications set in different domains and formulate a business problem for each application that can be solved with ML. We then carry out the complete workflow of integrating an ML solution on top of each application, \change{collecting} previously defined metrics to the code base at \change{each stage} of the deployment. Finally, we compare the fitness of each of FBP and SOA paradigm for ML deployment based on the measurements taken. Results show that FBP is able to address some of the key challenges around ML deployment by exposing the data in the system. At the same time there are still some gaps that remain before it can be considered a go-to paradigm for DOA. Specifically, our paper makes following contributions:
\begin{itemize}
    \item We quantitatively evaluate FBP in the context of ML deployment, analyzing its fitness with regard to data collection and model integration stages.
    \item We compare FBP with SOA, highlighting strengths of the former paradigm as well as identifying gaps that shall be covered to improve development experience with FBP.
    \item We show how FBP can be used to implement software systems according to DOA principles, by providing implementations of four applications set in different business domains.
\end{itemize}

\section{Motivation}\label{section:motivation}

Deployment of ML in production faces a variety of challenges that affect every step of the workflow \cite{paleyes2020challenges}. Data discovery and collection is one of the primary areas of concern \cite{polyzotis2018data, Lwakatare2020LargescaleML}.  In this section we explain how the problems in the currently most  prevalent software design approach motivated us to consider FBP.

Modern systems are built around "service-client" relationship\footnote{By service, or software service, we understand a software functionality or a set of software functionalities provided via an interface also known as an API, that different clients can reuse for different purposes by calling the named API.}. If a software client, perhaps a utility or a service or an object, needs an output of a certain operation, and there is a service capable of performing this operation, the client makes a call to that service passing in some input data through the service's API, and receives back a reply with the output data. An architectural paradigm that that is built entirely on this type of relationships is known service-oriented architecture (SOA) and is considered a de-facto standard of modern software engineering \cite{oreilly2020microservices}.

Such approaches that rely on API calls can often make data ephemeral, as we show later. That hinders efforts to re-use this data, for example in model training or validation. This presents a critical challenge for systems that are focused on data-driven decision making or data analysis.

\begin{figure}[h]
	\centering
	\includegraphics[width=0.6\columnwidth]{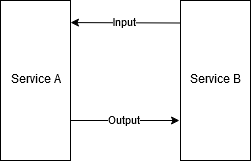}
	\caption{System of two software services. Service B sends a request with some input to service A, and receives a response with some output. Input and output data are passed over the network, and are not necessarily stored anywhere.}
	\label{figure:two_services}
\end{figure}

To illustrate why this is the case, let us consider a system of two services, A and B, where B is making a call to A, A does some computations internally, and then returns output to B, as shown on Figure \ref{figure:two_services}. Imagine we would like to collect a dataset that allows us to study the behavior of service A. That could be for the purposes of measuring its performance, running a business analysis task, or training a machine learning model. For that we would need a set of $(X, y)$ pairs, where each $X$ is a sample of an input that B sends to A, and each $y$ is the corresponding output that A sent back to B. In the current architecture there is nothing that guarantees such pairs were ever recorded. Even more so, there is no guarantee that $X$ or $y$ are available anywhere as separate data collection.

\begin{figure}[h]
	\centering
	\includegraphics[width=\columnwidth]{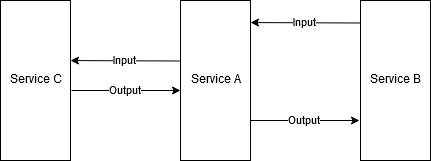}
	\caption{System of three software services. Service A makes an interim call to service C before generating a response to B's request. All inputs and outputs are passed over the network, and without separate effort it is difficult to match response from C to the corresponding request from B.}
	\label{figure:three_services}
\end{figure}

Consider also a more complex system with one more service C (Figure \ref{figure:three_services}). In this system, before replying to service B, service A also makes an interim call to service C to collect some additional input it needs to complete the computation. Here service A introduces hidden state, which complicates the task of collecting all the data required to describe A's behavior. Not only do we need to store and match inputs from B to the outputs, we also need to match each such pair to corresponding call to C. The problem quickly grows with the number of software services involved in the system.

In practice some of the necessary data can be found in logs and databases. However the decision on what data to store is left to developers of services A, B and C. Moreover, developers of different services within one software system may choose different technologies and formats for their logs and databases. In Confluent, the software company behind Apache Kafka, this situation is known as "The Data Dichotomy": while high quality data management is about exposing data, services are about hiding it \cite{stopford2016data}.

When data scientists are presented with a business problem, their first step is to understand what data is available within the system, and to collect it into a dataset suitable for ML model training. With the majority of data hidden behind services, data scientists have to spend significant time discovering this data. This involves talking to service owners, examining databases, parsing logs, merging multiple existing data sources, and partnering with software engineering teams to collect additional data. These struggles are described in detail by Lin and Ryaboy, who describe how the scalable service-oriented architecture in Twitter becomes a source of multiple problems for data science projects inside the company \cite{lin2013scaling}. Authors point out that while each service outputs rich logging information which could, in theory, serve as a dataset source, in practice these logs are usually inconsistent, incomplete and difficult to parse. Furthermore, the single responsibility principle that drives SOA architectures means it is difficult for data scientists to collect a complete dataset, and many data science projects get stuck at the data discovery stage. Similar observations are made by Nazabal et al.~\cite{nazabal2020data}, who collectively call these issues `data organization' and recognize them as one of the major challenges data scientists have to solve.

The main point is that there is nothing embedded in the SOA design paradigm that simplifies data discovery and collection. Any data-driven tasks within the boundaries of this system requires separate efforts from service developers and data scientists. \change{Organizational measures, such as team structures and conventions, can address this to a certain extent. However organizational measures on their own scale poorly and tend to break in long term, and shall be supported by strong and agile architectural framework \cite{Larsson2007AssessingTI, Nord2014AgileID}}. This motivated our search for an alternative architecture paradigm that considers data a first-class citizen of a software system, thus fulfilling the growing demand for creating data-driven solutions.

\section{Background}\label{section:background}

In this section we briefly introduce concepts and paradigms used throughout the paper.

\subsection{Data-oriented architecture}

Data-oriented architecture (DOA) is an emerging software pattern, where the goal is to create real-time information systems without centralised servers~\cite{vorhemus2017data}. DOA proposes to treat the system's data as a first class citizen in a shared information model with its respective access methods. System components in DOA are distributed, stateless, and communicate between each other using an asynchronous message exchange protocols~\cite{joshi2007data}. Such characteristics have the potential to support software developers when addressing ML deployment challenges~\cite{paleyes2020challenges}. Early ideas of applying DOA in ML deployment context were first introduced by Diethe et al.~\cite{diethe2019continual} and further developed by Lawrence~\cite{lawrence2019doa} and Borchert~\cite{borchert2020milan}. In these works DOA is described as a design approach that uses data streaming to create data-first software systems. Borchert also presented Milan, a domain-specific language that can be used for building systems according to the DOA principles. However, the most practical way to implement DOA in practice is an unresolved question, as references above all employ different methods and do not provide a clear comparison with modern software engineering practices. In that light our work can be seen as a step towards moving DOA beyond a high level concept by applying it to practical tasks and comparing against other popular software design methodologies.

\subsection{Flow-based programming}

Flow-based programming was created by J.P. Morrison \cite{morrison1994flow}, and can be considered a special case of the more general dataflow programming paradigm. It realises the DOA principles as it defines software applications as a set of processes which exchange data via connections that are external to those processes. FBP exhibits `data coupling', which is considered in computing to be the loosest form of coupling between components, thus promoting a flexible software architecture. FBP has a reputation as a paradigm that optimizes speed, bandwidth and processing power in multi-tasking, parallel processing applications. For example Szydlo et. al. consider FBP's application to IoT \cite{szydlo2017flow}, Lampa et.al. explored FBP's potential in the context of drug discovery \cite{lampa2016towards}, Zaman et. al. presented an FBP-based framework for mobile development \cite{zaman2015flow}. Recent years saw birth of several general-purpose projects built on flow-based programming principles, such as NoFlo \cite{bergius2015noflo} and Node-RED\footnote{\url{https://nodered.org/}}. Node-RED in particular became popular in the IoT community \cite{clerissi2018towards, chaczko2017learning}, as the FBP model was found to be a good fit for building data processing pipelines in IoT. Developing this idea further, in this paper we argue that there is potential in a wider use of FBP beyond IoT.

A notable feature of FBP is the ability to present the whole program visually as a graph of data flow. This feature has two important implications. First, the graph-like structure allows reasoning about the entire system in a way that is often impossible in case of OOP or SOA. We leverage this aspect later in this work. Second, it allows for visual no-code programming which in some cases may aid adoption of FBP. In particular it is believed to be useful for beginners who have little or no prior coding experience \cite{mason2017block}.

We make use of data streams as connectors in flow-based programs. A data stream is a sequence of data records that are made available over time. Machine learning on data streams is not a new concept. Data processing platforms such as Apache Spark \cite{meng2016mllib}, Apache Flink \cite{carbone2015apache} or Google Cloud Dataflow \cite{krishnan2015google} are widely used for manipulating large datasets and executing machine learning tasks. Apache Kafka \cite{kreps2011kafka} and AWS Kinesis\footnote{\url{https://aws.amazon.com/kinesis/}} are some of the most commonly used data streaming services.

\subsection{Service-oriented architecture}

Service-oriented architecture (SOA) is a well known paradigm for development of software applications \cite{perrey2003service}. Under this paradigm the application is broken down into several components called services, which interact between each other via a pre-defined protocol to provide the application's users with necessary functionality. Services only interact on API level, thus hiding details of each service's implementation, which results in a set of loosely coupled components. Service orientation gives developers a range of important benefits: encapsulation, loose coupling, modularity, scalability and ease of integration with other systems, so it is a reasonable choice for those who need to build scalable software~\cite{papazoglou2003service,cabrera2017implementing}. In recent years an SOA derivative known as microservices gained substantial popularity and can be considered de-facto standard distributed software design approach \cite{oreilly2020microservices}. \change{Therefore for the reminder of this paper we will assume microservices approach when talking about SOA.}
\begin{table*}[!ht]
\caption{Evaluation Metrics}
\resizebox{1.9999\columnwidth}{!}{
\label{tab:metrics}
\begin{tabular}{|l|l|l|}
\hline
\multicolumn{1}{|c|}{\textbf{Goals}}                                                             & \multicolumn{1}{c|}{\textbf{Questions}}                                                          & \multicolumn{1}{c|}{\textbf{Metrics}}                                                   \\ \hline
\multirow{4}{*}{
\begin{tabular}[c]{@{}l@{}}Goal 1: Analysing software paradigms \\ for the purpose of evaluating their \\ impact in the software development\\ process with respect to \textbf{data collection tasks} \\ from the point of view of \\ software developers in the context of\\ the development of data processing \\ applications.
\end{tabular}}           & 
\begin{tabular}[c]{@{}l@{}}Question 1: How much additional code is required \\ to implement a data collection task in the evaluated \\ applications for each paradigm?
\end{tabular}                 & 
Logical lines of code                                                         \\ \cline{2-3} & \begin{tabular}[c]{@{}l@{}}Question 2: How does the system's maintainability \\ change when data collection tasks are implemented \\ in the evaluated application for each paradigm?
\end{tabular}   & 
Maintainability Index                                                           \\ \cline{2-3} &
\begin{tabular}[c]{@{}l@{}}Question 3: How complex does the evaluated application \\ become after implementation of data collection task \\ for each paradigm?
\end{tabular}                 & 
\begin{tabular}[c]{@{}l@{}}Cognitive Complexity
\end{tabular} \\ \cline{2-3} &
\begin{tabular}[c]{@{}l@{}}Question 4: How intrusive is the dataset collection \\ task in the evaluated applications for each paradigm?
\end{tabular}                                                 & 
Number of affected components                                                                      \\ \hline
\multirow{4}{*}{\begin{tabular}[c]{@{}l@{}}Goal 2: Analysing software paradigms\\ for the purpose of evaluating their \\ impact in the software development\\ process with respect to \\ \textbf{ML model integration} \\ from the point of view of software developers \\ in the context of the development of \\ data processing applications.
\end{tabular}} & 
\begin{tabular}[c]{@{}l@{}}Question 1: How much additional code is required \\ to implement an integration task in the evaluated \\ applications for each paradigm?
\end{tabular}                      & 
Logical lines of code                                                         \\ \cline{2-3} &
\begin{tabular}[c]{@{}l@{}}Question 2: How does the system's maintainability \\ change when model integration task is implemented \\ in the evaluated applications for each paradigm?
\end{tabular} & 
Maintainability Index                                                           \\ \cline{2-3} &
\begin{tabular}[c]{@{}l@{}}Question 3: How complex does the evaluated application \\ become after integration of ML model for each paradigm?
\end{tabular}                                  & 
\begin{tabular}[c]{@{}l@{}}Cognitive Complexity
\end{tabular} \\ \cline{2-3} &
\begin{tabular}[c]{@{}l@{}}Question 4: How intrusive is model integration task\\ in the evaluated applications for each paradigm?
\end{tabular}                                                       & 
Number of affected components                                                                      \\ \hline
\end{tabular}
}
\end{table*}

\section{Experiment design and implementation}\label{section:design}

Our evaluation of the DOA-based applications follows the methodology proposed by Wohlin et. al.~\cite{wohlin2012experimentation}. We first define a set of metrics following the GQM framework, and then develop four applications that cover a variety of business domains and areas of ML\makeatletter
\footnote{
\if@ACM@anonymous
   Implementation of all applications is openly available on GitHub. Link to the repository is removed for blind review.
\else
   Full source code of the project can be found at \url{https://github.com/mlatcl/fbp-vs-soa}
\fi}
\makeatother.
These applications allow us to study the properties of FBP and compare it against classical SOA approach in the experimental setting. In this section we describe the metrics we used, the applications we implemented, and give details of the experiment we defined to examine fitness of both paradigms in the ML deployment context.

\subsection{Metrics Definition}
We follow the GQM framework to define the metrics evaluation \cite{wohlin2012experimentation}. This framework proposes to define the evaluation goals as a first step. These goals are then mapped to metrics through questions required to achieve the goals. Table~\ref{tab:metrics} introduces the goals of this evaluation, the questions we want to answer to achieve these goals, and the metrics that will allow us to provide such answers. The resulting metrics are defined as follows:

\begin{itemize}
    \item \textbf{Logical Lines of Code}, which counts executable statements and ignores comments and blank lines. We use it as a measure of an application's codebase size. Specifically, we assess how much additional code is needed to implement additional functionalities in the applications.
    \item \textbf{Maintainability Index}, as defined in the Radon package\footnote{\url{https://radon.readthedocs.io/en/latest/intro.html\#maintainability-index}}, is a composite metric that is calculated using a number of other metrics as operands. This is a unitless metric that assigns a codebase a score between 0 and 100. We use it to assess how maintainable is the application's codebase. 
    \item \textbf{Cognitive complexity} \cite{campbell2018cognitive}, which measures the complexity of the control flow of code. We use it to assess how easy it is to understand the applications code. This metric is similar to McCabe's cyclomatic complexity \cite{mccabe1976complexity}, and was proposed as a replacement that focuses on human's understanding of the source code, which is critical for software development and maintenance. Since cognitive complexity is measured separately for each code block, we consider average cognitive complexity across the whole codebase.
    \item \textbf{Number of Affected Components}, which counts the number of components that were added or changed during a certain stage of development. It allows us to evaluate intrusiveness of a particular feature, that is how many parts of the codebase had to be changed or added for implementation of that feature. For the purposes of this experiment we identify processing nodes and data streams as components of a FBP program, and APIs and data access routines as components of SOA program.
\end{itemize}

\subsection{Applications}

We have implemented four applications separately with FBP and SOA paradigms. For each application we formulated a business task that is illustrious of a typical task a practicing data scientist might face. We then carried out deployment of a data-powered solution to the task while observing the evolution of the codebase throughout the deployment cycle. In this section we give a brief description of each application as well as deployment stages.

\textbf{Ride Allocation} application maps incoming ride requests to available drivers and tracks history of existing rides. We have formulated a task of estimating pickup wait time for each ride allocation based on historical data. This task is approached as a supervised learning regression problem \change{with offline model training of the collected dataset. More elaborate description of this application and its implementation details can be found in our earlier work \cite{paleyes2021towards}.}

\textbf{MBlogger} is a micro-blogging platform that keeps track of users, their following/follower relationships between each other, and builds a timeline of posts for each user based on activity of those this user follows. As a task we decided to build a post-generating bot that, given a particular user, can generate posts this user is likely to be interested in. The solution can be considered a simple generative NLP approach. In MBlogger we do not collect offline dataset file, instead storing all data that is needed to generate posts \change{on the fly} using runtime infrastructure.

\textbf{Insurance Claims} models a workflow that processes insurance claims. Claims undergo a series of classification routines which affect the choice of the final payout process. In this application we use ML to replace all internal logic with one classification model. \change{As is the case with Ride Allocation, this model is trained offline and then is used for online inference.} Additionally, this application is different from the ones described above because albeit ML model is being deployed, it does not affect the user interface and only changes the internal data processing mechanism of the application.

\textbf{Playlist Builder}\footnote{This application closely follows Metaflow tutorial: \url{https://docs.metaflow.org/getting-started/tutorials/season-1-the-local-experience/episode01}} creates a movie playlist for a specified genre. Initial functionality builds playlists at random, and later stages only add highest-grossing movies to a playlist. Unlike all other applications, here we do not collect raw data or train any models. Instead quantiles of movies' gross earnings are calculated, and then used for filtering \change{online}. Playlist Builder illustrates a simple yet realistic use case where a data scientist \change{needs to build a solution that collects certain statistics about the data and then makes automated decisions based on it}.

We defined three stages of the implementation of each application to evaluate codebase changes:
\begin{itemize}
    \item \textbf{Stage 1}: minimal code to provide basic functionality without any ML-powered capabilities. The stage is denoted in the code and this paper by suffix \textit{min}.
    \item \textbf{Stage 2}: code for Stage 1 plus implementation of data collection. Denoted by suffix \textit{data}.
    \item \textbf{Stage 3}: code for Stage 2 plus code necessary for integrating the data-driven solution. If the ML model is created, it is trained on the dataset collected at the previous stage. Denoted by suffix \textit{ml}.
\end{itemize}

Our experiment contains six implementations of each of the applications described above. Table \ref{table:version_list} summarizes these versions, and Figure~\ref{figure:all_stages} illustrates all six implementations for one of the applications (Insurance Claims) with sequence (for SOA) and data flow (for FBP) diagrams.

\begin{table*}[t]

	\caption{List of all created versions created for each application. First column gives the key by which a particular version is referred to in the codebase and in the paper.}
	\label{table:version_list}
	\begin{center}
		\begin{tabular}{ |l|l|l|l| }
			\hline
			\textbf{Key} & \textbf{Paradigm} & \textbf{Stage} & \textbf{Description} \\
			\hline
			$fb\_app\_min$ & FBP & 1 & Basic functionality \\
			\hline
			$fb\_app\_data$ & FBP & 2 & Same as $fb\_app\_min$ plus data collection \\
			\hline
			$fb\_app\_ml$ & FBP & 3 & Same as $fb\_app\_data$ plus model integration \\
			\hline
			$soa\_app\_min$ & SOA & 1 & Basic functionality \\
			\hline
			$soa\_app\_data$ & SOA & 2 & Same as $soa\_app\_min$ plus dataset collection \\
			\hline
			$soa\_app\_ml$ & SOA & 3 & Same as $soa\_app\_data$ plus model integration \\
			\hline
		\end{tabular}
	\end{center}

\end{table*}

\subsection{Implementation notes}
Each application in the experiment is structured the same way: the application itself and the code that simulates events happening in the outside world. The application part processes the incoming data according to some business logic, and outputs data that is then used by the simulation. The simulation part is implemented as a discrete-event simulation and is responsible for generating events that would be happening if the system was deployed in real life.

The SOA version of each application is done in a form of a RESTful service. We use the lightweight Flask\footnote{Flask Framework - \url{https://palletsprojects.com/p/flask/}} framework to develop the SOA applications because of its flexibility and popularity among the Python community. Microservices persist and manipulate data using SQLite\footnote{SQLite - \url{https://www.sqlite.org/index.html}}. This popular database engine offers a small, fast, self-contained, high-reliability, full-featured platform for data management. We implement a set of services that offer the required capabilities for each application and implementation stage, and data access layer that manages database queries. All services are hosted locally, and the communication is happening via HTTP requests.

The FBP version of each application is built with two major building blocks: data streams and stateless processing nodes. We use the lightweight Python "FBP-inspired" framework flowpipe\footnote{flowpipe - \url{https://github.com/PaulSchweizer/flowpipe}}. A data stream is a collection of data records of the same type that is updated whenever a new record arrives. Each data stream within the application belongs to one of three categories:
\begin{itemize}
	\item Input streams, that receive data from the outside world;
	\item Output streams, that hold data produced by the application;
	\item Internal streams, that hold intermediate data within the application. These streams are necessary because processing nodes by definition are not allowed to have state.
\end{itemize}

A processing node takes one or more streams as an input, performs some operations on them, and then puts the result into output or internal data streams. Such nodes do not carry internal state, and do not make external calls to outside services or databases. All data influencing a processing node should be registered in the system, so if such additional input is necessary, it should be represented as a data stream.

\begin{figure*}[p]
	\newcommand{\scale}{1.0}
	\begin{subfigure}[]{\scale\columnwidth}
	    \centering
		\includegraphics[width=\columnwidth]{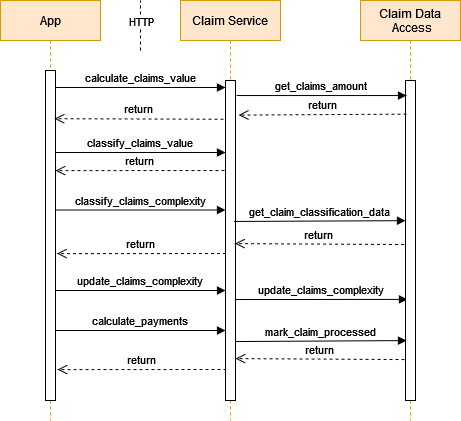}
		\caption{SOA \textit{min} stage.}
	\end{subfigure}
	\begin{subfigure}[]{\scale\columnwidth}
		\centering
		\includegraphics[width=\columnwidth]{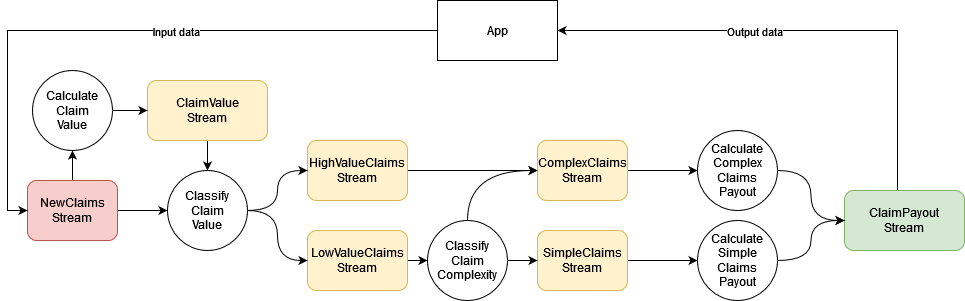}
		\caption{FBP \textit{min} stage.}
	\end{subfigure}
	\begin{subfigure}[]{\scale\columnwidth}
		\centering
		\includegraphics[width=\columnwidth]{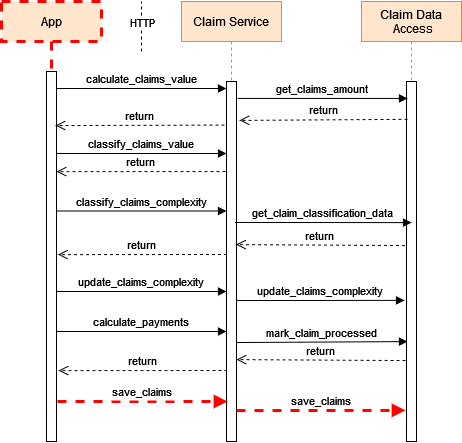}
		\caption{SOA \textit{data} stage.}
	\end{subfigure}
	\begin{subfigure}[]{\scale\columnwidth}
		\centering
		\includegraphics[width=\columnwidth]{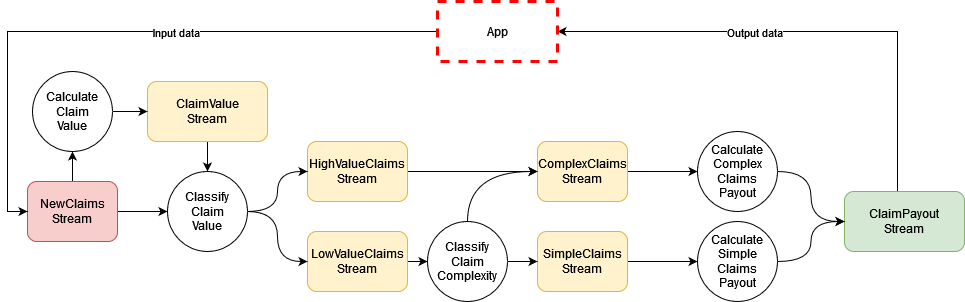}
		\caption{FBP \textit{data} stage.}
	\end{subfigure}
	\begin{subfigure}[]{\scale\columnwidth}
		\centering
		\includegraphics[width=\columnwidth]{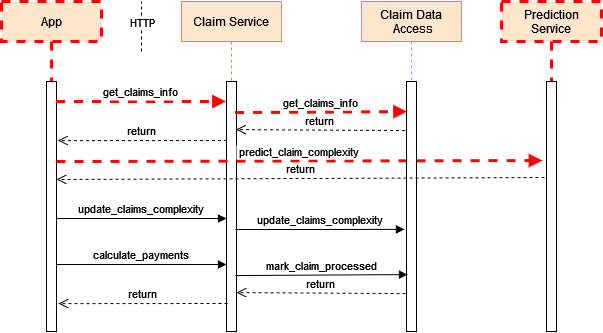}
		\caption{SOA \textit{ml} stage.}
	\end{subfigure}
	\begin{subfigure}[]{\scale\columnwidth}
		\centering
		\includegraphics[width=\columnwidth]{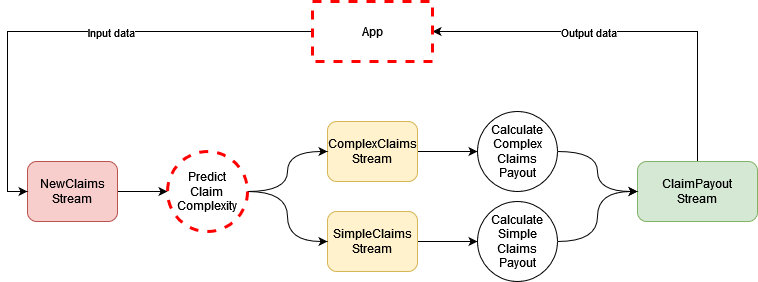}
		\caption{FBP \textit{ml} stage.}
	\end{subfigure}
	\caption{Diagrams of Insurance Claims app evolution through three stages of ML deployment for both paradigms. We use sequence diagrams for SOA (left) and data flow diagrams for FBP (right). \change{Yellow boxes in SOA diagrams represent individual services. Circles in FBP diagrams represent processing nodes, and boxes represent data streams: red for input, yellow for internal, green for output streams.} Deployment stages top to bottom are: \textit{min, data, ml}. New or updated components and APIs at each subsequent stage are highlighted with bold red dashed lines.}
	\label{figure:all_stages}
\end{figure*}

Figure~\ref{figure:all_stages} shows diagrams that describe all 6 versions of one app\footnote{We have chosen to show only one application, Insurance Claims, because of space constraints.}. As can be seen on these diagrams, entry point for all versions is an App object, that either orchestrates calls to necessary services (SOA) or triggers an evaluation of the data flow graph (FBP). Within each paradigm subsequent stages introduced new APIs or graph nodes, all of which are highlighted for clarity.

Since applications are implemented using Python frameworks, we used Python tools Radon\footnote{Radon - \url{https://radon.readthedocs.io/en/latest/}} and Flake8\footnote{Flake8 - \url{https://flake8.pycqa.org/en/latest/}} for metrics collection.
\section{Evaluation}\label{section:evaluation}
In this section we present the results of the experiment. We follow the goals and questions formulated in Table~\ref{tab:metrics} and answer each question by analyzing the corresponding metric.

\subsection{Data collection task}
First we address the questions around the data collection task, by measuring the changes between \textit{min} and \textit{data} stages.

\begin{figure*}[t]
    \newcommand{\scale}{0.9}
	\begin{subfigure}[b]{\scale\columnwidth}
		\includegraphics[width=\columnwidth]{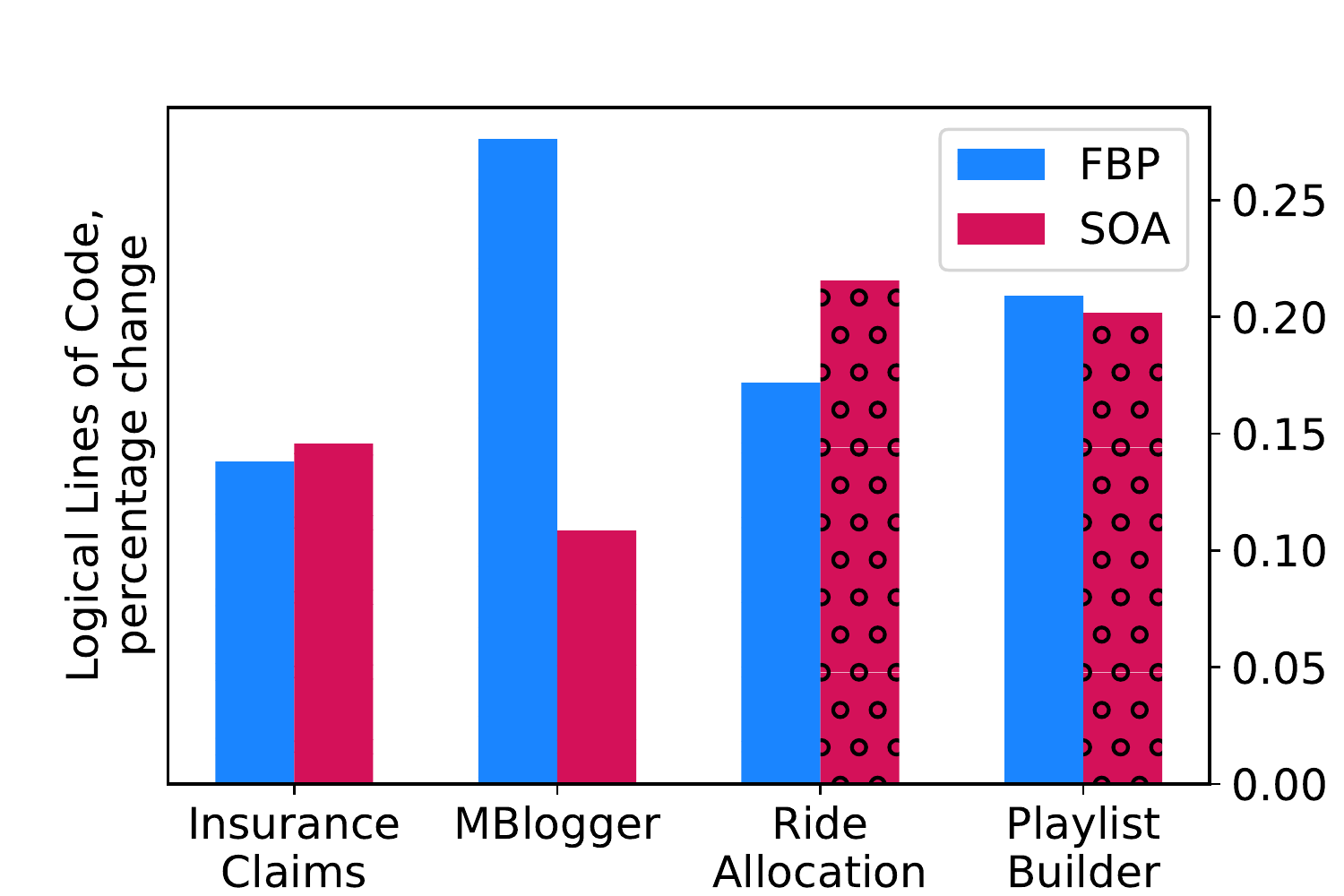}
		\caption{}
		\label{figure:lloc_min_data_diff}
	\end{subfigure}
	\begin{subfigure}[b]{\scale\columnwidth}
		\includegraphics[width=\columnwidth]{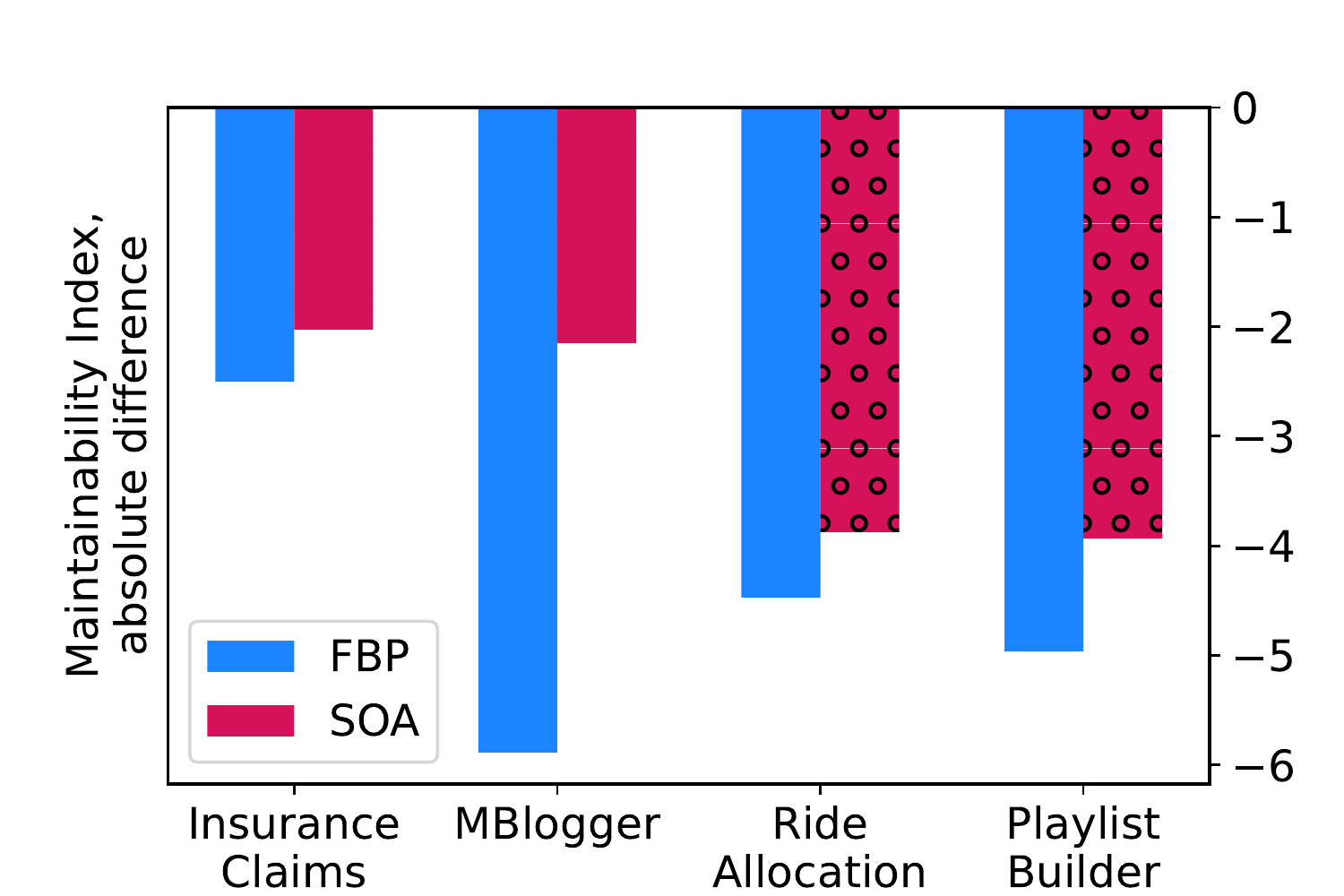}
		\caption{}
		\label{figure:mi_min_data_diff}
	\end{subfigure}
	\begin{subfigure}[b]{\scale\columnwidth}
		\includegraphics[width=\columnwidth]{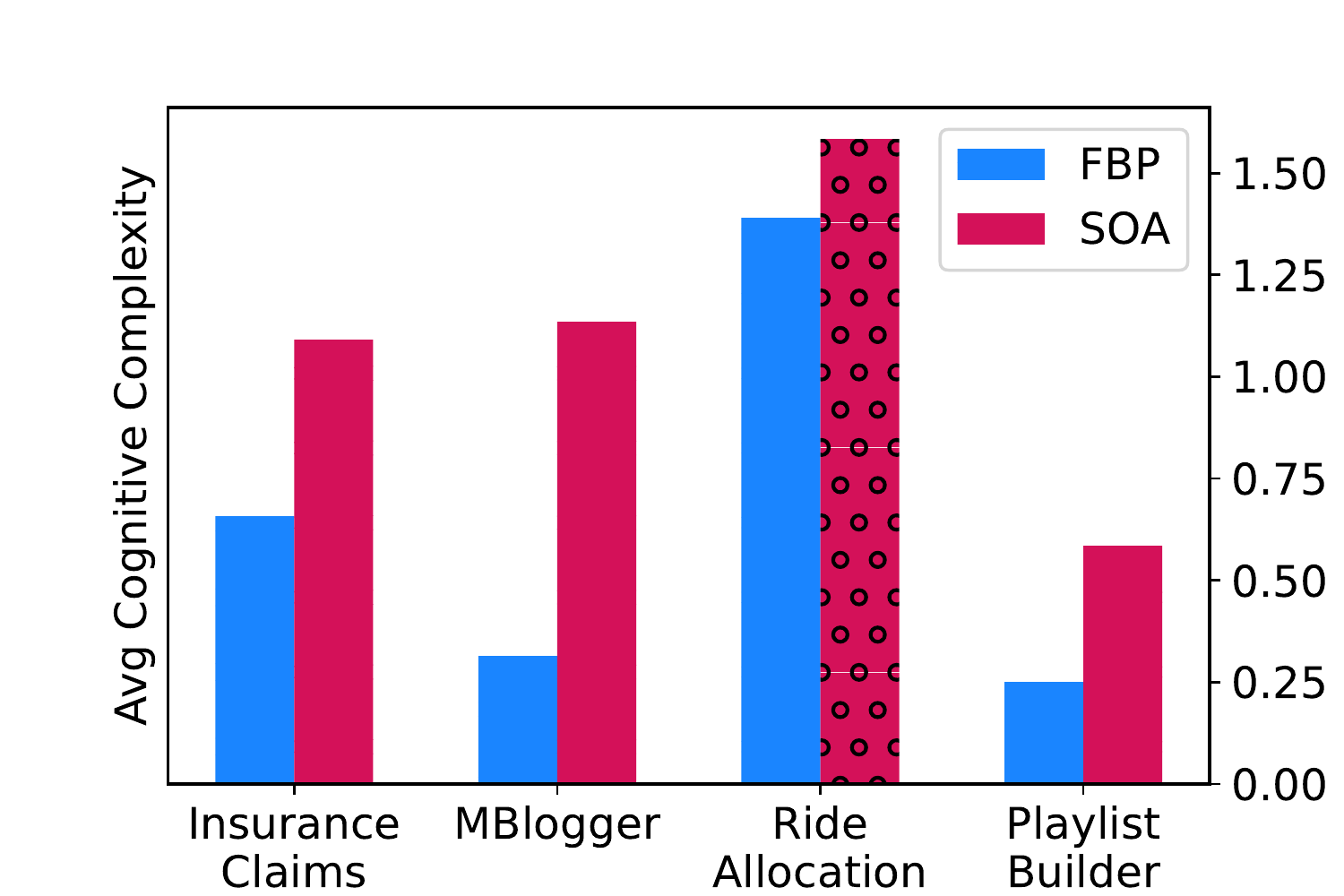}
		\caption{}
		\label{figure:cognitive_data}
	\end{subfigure}
	\begin{subfigure}[b]{\scale\columnwidth}
		\includegraphics[width=\columnwidth]{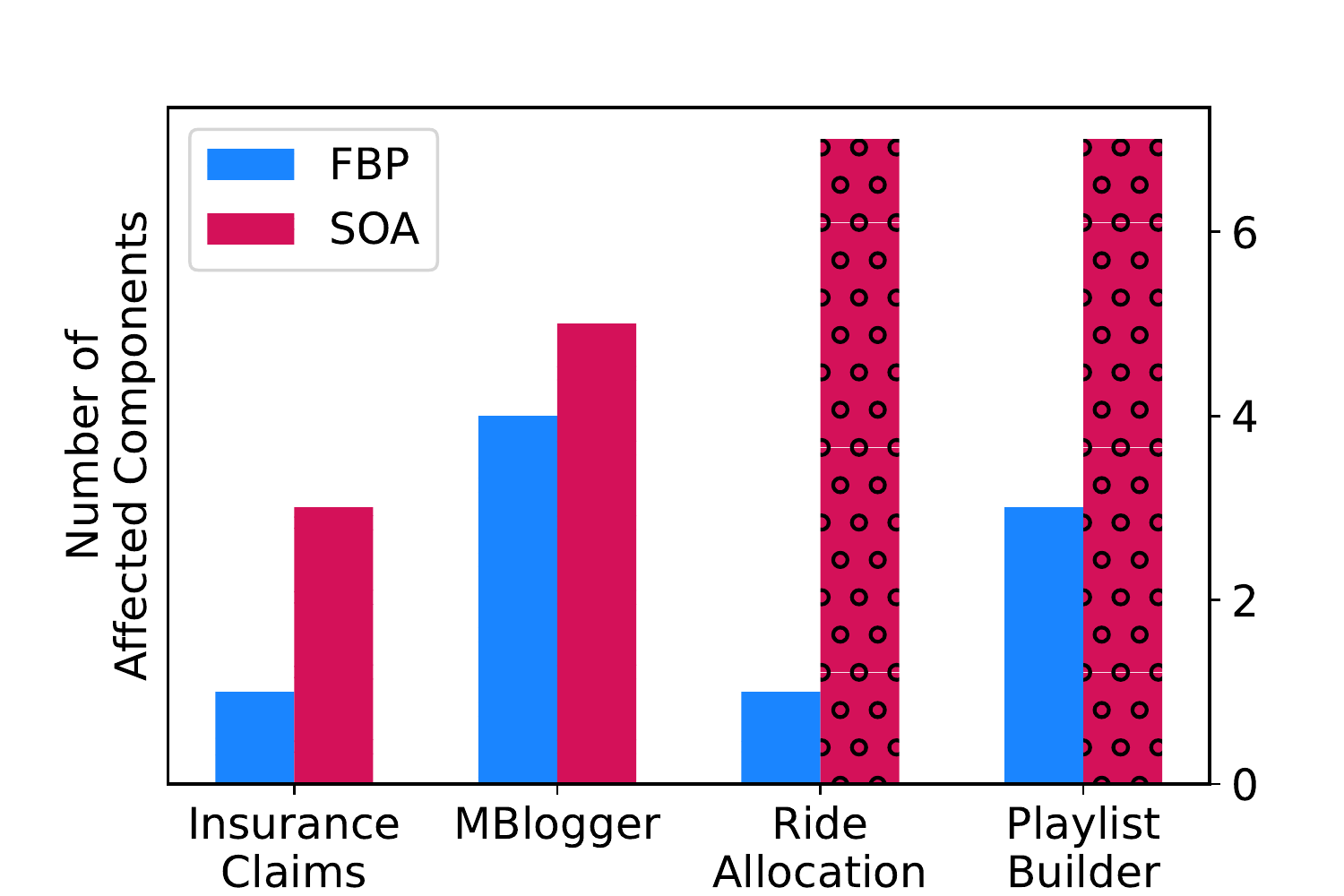}
		\caption{}
		\label{figure:components_min_data_diff}
	\end{subfigure}
	\label{figure:min-data-metrics-diff}
	\caption{Measurements of the impact data collection task had on FBP and SOA implementations of four applications.}
\end{figure*}

\subsubsection{How much additional code is required to implement a data collection task?}

We used Logical Lines of Code (LLOC) metric to answer this question, thus measuring how much code, in percentage of the initial size of the codebase, was added for the data collection stage. We use percentage change to make this comparison independent of the amount of boilerplate code that might be different for different paradigms. As can be seen from Figure~\ref{figure:lloc_min_data_diff}, two out of four FBP and SOA applications required comparable amount of additional code to implement data collection. In both cases the difference is small: growth difference was less than 0.2\% for Insurance Claims codebase and 4\% for Ride Allocation codebase.  MBlogger and Playlist Builder show different behavior, suggesting that setting up additional runtime infrastructure for data may bear higher initial cost for FBP\footnote{As a reminder, MBlogger and Playlist Builder do not create an offline dataset file to accomplish the data collection task.}. Overall SOA exhibits same or slower growth of the codebase for implementation of data collection.

\subsubsection{How does the system’s maintainability change when data collection tasks are implemented?}

We used Maintainability index (MI) metric to answer this question. MI is a score from 0 to 100, and does not scale with the size of the codebase, which allow us to compare its absolute values. For three out of four applications maintainability was not dramatically impacted in either paradigm, dropping similar amount of score points (between 2 and 5). But importantly, MI score has consistently decreased more for the FBP codebase than it did for the SOA codebase across all four applications, as can be seen on Figure~\ref{figure:mi_min_data_diff}. The reason for that is the fact that to collect data in FBP we needed to implement a relatively complex dataflow graph traversal logic. Since all our codebases are small (average size is 314 LLOC), this traversal code had a great impact on the MI metric. A tool designed to collect data from a dataflow graph that encapsulates such traversal operations would improve developer experience.

\subsubsection{How complex does the application become after implementation of the data collection task?}

This question is answered with the help of cognitive complexity metric. To make this comparison independent of the size of the codebase, we measure complexity of each block of code independently, and average these measurements across the whole codebase. Results can be found on Figure \ref{figure:cognitive_data}. Because of its focus on modeling how the data flows through the system, FBP code shows lower average complexity across all four applications, and on three occasions the cognitive complexity is at least twice as low. This suggests that overall system is simpler to comprehend when it models dataflow, and thus might be easier work with in long term.

\subsubsection{How intrusive is the data collection task?}

We counted the number of components that were affected to implement the data collection task. FBP required less changes for all applications (Figure~\ref{figure:components_min_data_diff}). Only one component had to change for two out of four FBP codebases that collected the dataset for offline. Lower impact of FBP in this metric can be explained with the ability of FBP programs to traverse the dataflow graph, which allows for more localized changes in the system when it comes to data collection.

\subsection{ML model integration task}
We now address the same questions in the context of ML model integration, by comparing measurements of \textit{data} and \textit{ml} stages.

\begin{figure*}[t]
	\newcommand{\scale}{0.9}
	\begin{subfigure}[b]{\scale\columnwidth}
		\includegraphics[width=\columnwidth]{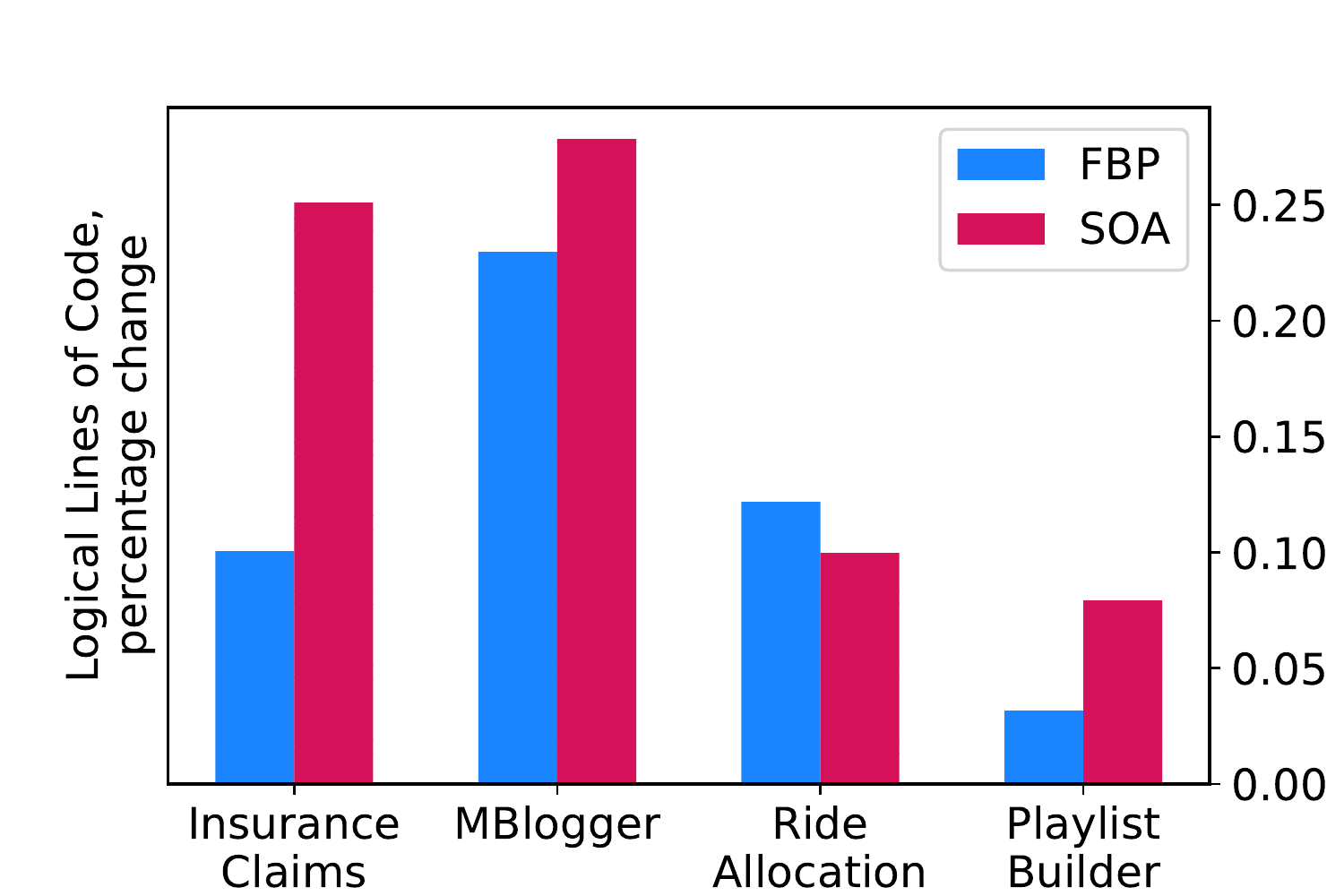}
		\caption{}
		\label{figure:lloc_data_ml_diff}
	\end{subfigure}
	\begin{subfigure}[b]{\scale\columnwidth}
		\includegraphics[width=\columnwidth]{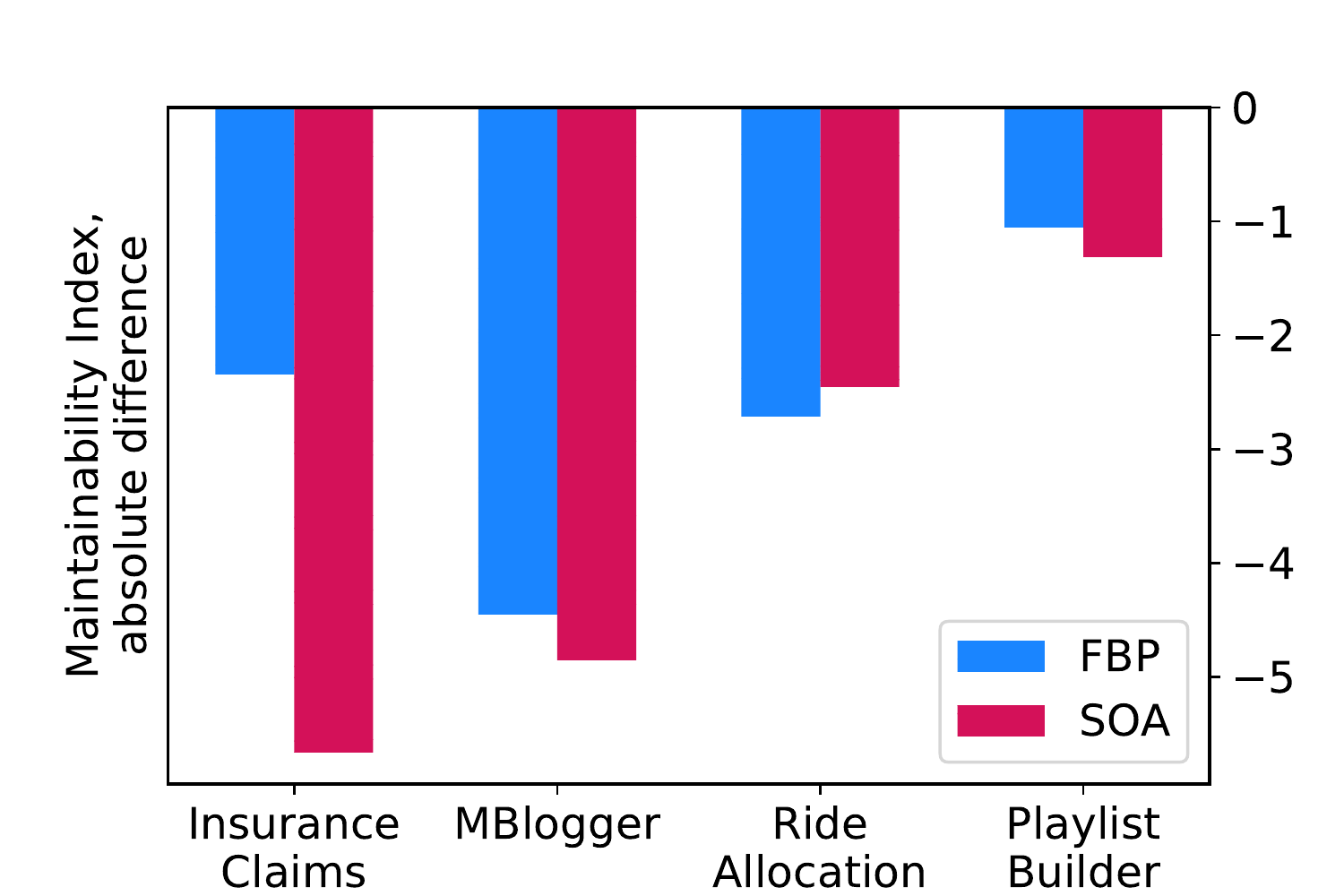}
		\caption{}
		\label{figure:mi_data_ml_diff}
	\end{subfigure}
	\begin{subfigure}[b]{\scale\columnwidth}
		\includegraphics[width=\columnwidth]{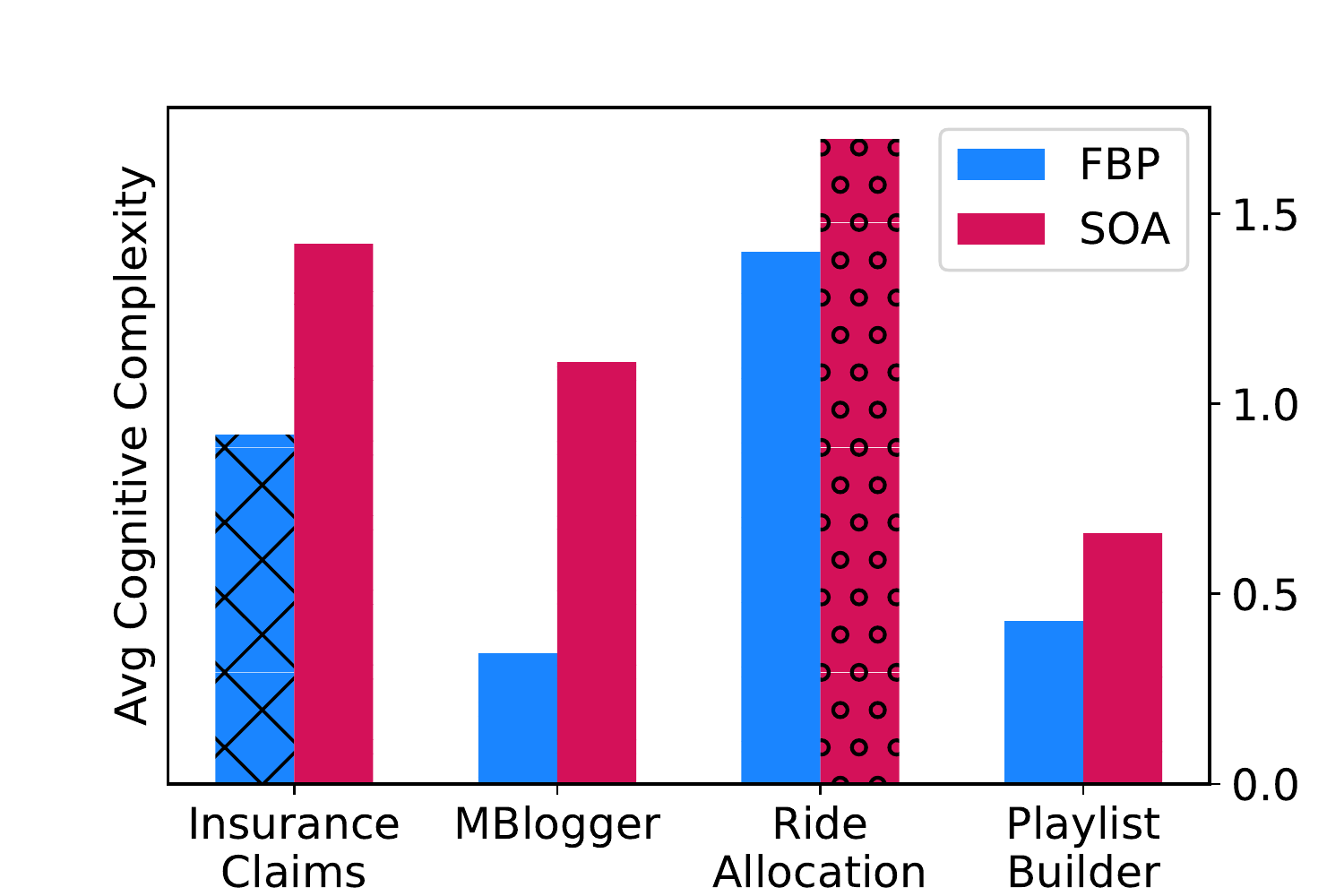}
		\caption{}
		\label{figure:cognitive_ml}
	\end{subfigure}
	\begin{subfigure}[b]{\scale\columnwidth}
		\includegraphics[width=\columnwidth]{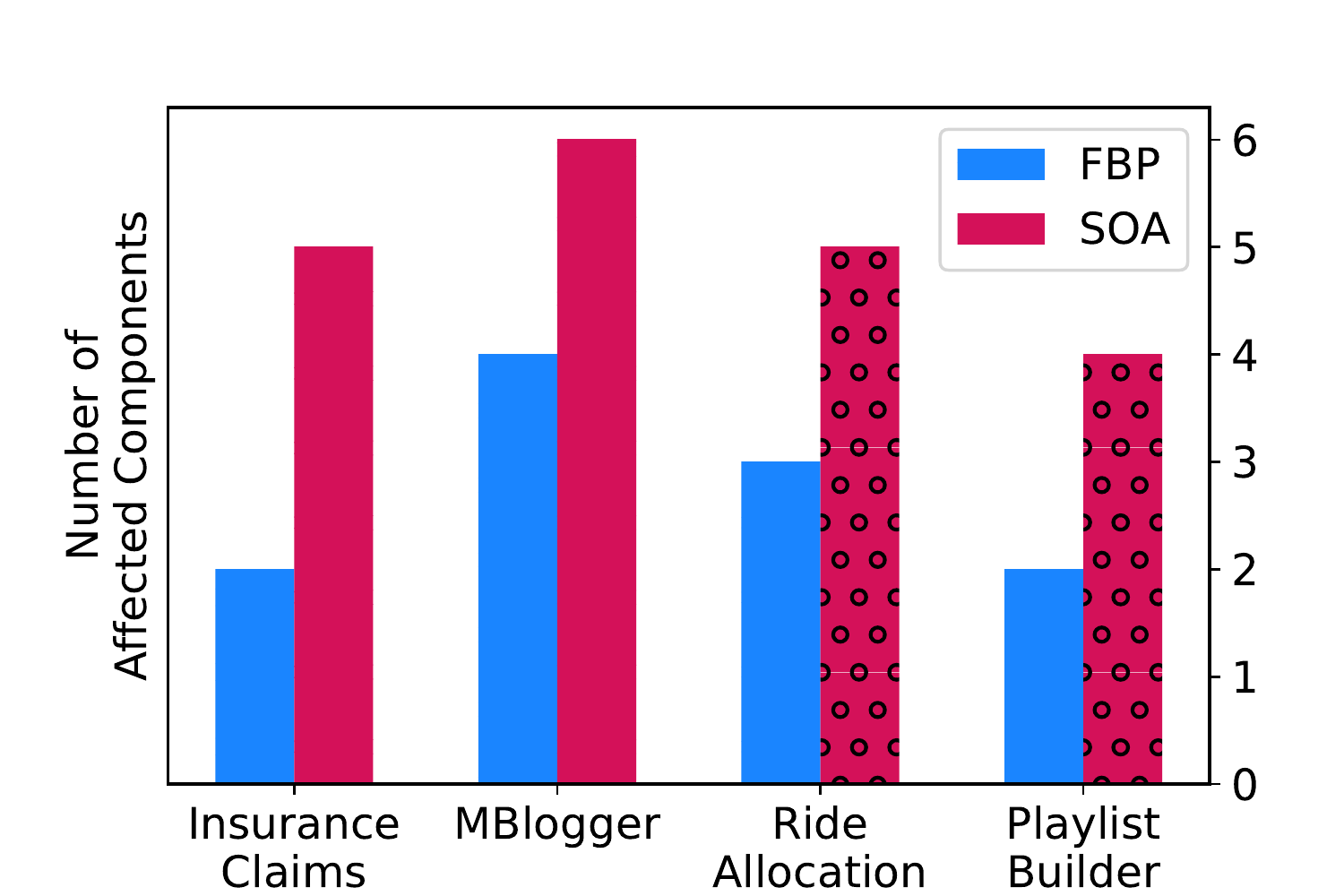}
		\caption{}
		\label{figure:components_data_ml_diff}
	\end{subfigure}
	\label{figure:data-ml-metrics-diff}
	\caption{Measurements of the impact ML model integration task had on FBP and SOA implementations of four applications.}
\end{figure*}

\subsubsection{How much additional code is required to implement the model integration task?}

FBP codebase grew relatively slower at model integration stage for three applications: growths difference is 15\% for Insurance Claims, 5.3\% for MBlogger and 10\% for Playlist Builder, as can be seen on Figure~\ref{figure:lloc_data_ml_diff}. The difference between the paradigms is pronounced than at the previous stage, which is a benefit of loose data coupling. We conclude that FBP requires same or less effort to host a trained model than SOA.

\subsubsection{How does the system’s maintainability change when the model integration task is implemented?}

In terms of maintainability, FBP exhibited similar behavior at this stage compared to data collection, dropping between 1 and 4 points, see Figure~\ref{figure:mi_data_ml_diff}. Interestingly, SOA exhibited a different behavior, with much more severe impact for Insurance Claims and MBlogger: MI of these applications decreased by 4 and 5 points, compared to 2 points for data collection. The maintainability of these applications is more impacted because they need more code (Figure~\ref{figure:lloc_data_ml_diff}) at this stage. FBP shows more maintainable way of adding a component that hosts a ML model.

\subsubsection{How complex does the application become after implementation of the ML model integration task?}

Both paradigms allow to host a trained models or other data-driven components in a way that is loosely coupled with the rest of the system. Therefore the picture around cognitive complexity does not change significantly when compared to data stage - FBP solutions are still easier to comprehend, and deployment of data-driven components does not affect the complexity of the overall solution in either paradigm.

\subsubsection{How intrusive is the model integration task?}

As was the case with the data collection task, FBP again required changes in fewer components in all four cases. But the underlying reason for this is different. Recall that in data collection stage FBP required a dataflow graph traversal, which on two occasions could be implemented from a single component of the application. For model hosting no traversal is required. However the model needs to be given the data to make predictions. In case of FBP that only means additional wiring within the graph, to make sure model node is connected to necessary data streams. In SOA that means changes that have to made both on service and data access layer, to ensure right services are invoked and right data is fetched for the model. All in all, FBP allows for less intrusive changes for both stages of the deployment.

\section{Discussion}\label{section:discussion}
We evaluated FBP as a potential paradigm for building DOA software because of the way it exposes data passing through the system via the dataflow graph. We expected FBP to be a clearly better choice for data management and ML deployment than SOA. In this section we discuss the results of our evaluation and compare the experiment outcome against these expectations.

SOA-based applications showed smaller size growth in data collection phase. In contrast, FBP codebase often has to deal with traversing the dataflow graph for the purposes of collecting an offline dataset from multiple sources, which can be a complex operation to implement. Additionally, FBP required more boilerplate code to add new components for those applications where data was collected online. All of these resulted in bigger impact on FBP codebases in LLOC and Maintainability Index metrics. However, size of the codebase is not the only factor when it comes to maintainability of a system, and FBP excelled in other aspects of maintainability. Changes in FBP code are always more localized and have to be made in fewer parts of the system (according to the Number of Affected Components metric). Smaller number of components' changes brings down the costs of maintenance and increases robustness of the system over time \cite{sommerville2011software}. FBP code is also easier to comprehend, according to the Cognitive Complexity metric, which is again beneficial for maintenance. Quantification of these benefits of DOA systems implemented with FBP can be a subject to a long-term observational follow-up study. 

Interestingly, FBP code showed same or less impact by the model integration stage across all four metrics. This result might be surprising considering that SOA is usually chosen by professionals because of its loose coupling of components, and extensibility and scalability benefits it brings. This kind of robustness to deployment of a data-driven component is the direct consequence of data coupling that DOA promotes and FBP exhibits via external data connections principle.

The evaluation highlighted the gaps that need to be covered to make FBP a default choice for building DOA software. We have kept the number of programming tools and frameworks to a minimum, aiming at comparison of the paradigms in their vanilla state. This also means our evaluation can highlight the areas where appropriate tools can make the biggest improvement for the overall development experience. For FBP this is clearly the processing of the dataflow graph. Operations on that graph add a lot of complexity (e.g., node wiring and traversals). We believe that targeted tools that provide such operations would simplify the development process with FBP and thus increase the adoption of this paradigm for data-driven applications.

DOA promotes explicit handling of all data in the system, which makes data discovery available by design. FBP realises this principle with the dataflow graph. As discussed in Section~\ref{section:motivation}, in the control flow approaches (e,g. SOA) data and data relationships are difficult to discover. This problem grows with scale, as data becomes scattered between multiple data storage facilities. Therefore the dataflow graph feature is a significant improvement in FBP applications compared to SOA. \change{Notably, ability to build, access and traverse this graph can be achieved with hybrid approaches as well, for instance utilizing FBP approach for definition of the whole graph and microservices or serverless computing for implementation of individual components.}
\section{Threats to Validity}
\subsection{Internal validity}
We used a particular toolset for development of all the applications described in this study. Specifically, we used Python for all our development, Flask for SOA implementations, and flowpipe for FBP implementations. The particular choice of tools might have affected the code metrics collected. We have compensated for this in a number of ways. First, we have kept the number of tools to a minimum, to ensure we evaluate the paradigms on their own, and implementing our own boilerplate whenever necessary (e.g. data stream abstraction for FBP, entity mapping for SOA). Second, we developed different types of applications from different domains following the same architectures and using the same tools. This variety allows us to counter the possible tools impact.

Code metrics we used might be affected by the particularities of development process within a single paradigm. For instance, size-dependent metrics can produce larger values for paradigms that require more boilerplate code. We accounted for that by considering relative growth (LLOC) or average values (Cognitive Complexity), thus converting the metrics to size-independent ones.

\subsection{External validity}
The main threat to external validity of our work is its environment. Ideally the experiment described in this paper shall be run in a real production setting and observed over multiple projects. However that would require running two production ready systems with equal functionality simultaneously. That would double the requirements on workforce and computational resources, which is prohibitively expensive to expect from a business. \change{Additionally, FBP at the moment is a niche paradigm, rarely used outside of IoT and embedded systems, which would introduce a skew into such study, either positive or negative depending on the business domain. In fact the main purpose of our work is a wider adoption of FBP and other dataflow-type approaches outside of these areas.} For that reason we designed the experiment to model the real ML deployment process as closely as possible, separating out jobs done by software engineers and data scientists, and considering a range of various data science tasks. \change{In that sense the scale of the experiment we chose is a trade-off between realistic setup and clarity of our message.} Nevertheless, long-term benefits of adopting FBP as a primary design paradigm for data driven projects are not fully explored in this paper, and can be a subject of an extension industrial study.

Design of software systems has to account for other aspects, besides availability and collection of data. Modern distributed systems have to be fast, resilient, scalable, and secure. We left evaluation of FBP in all these areas out of scope of this study to keep our work focused on data science related tasks. However, all of these aspects are important and shall influence design choices made by software engineers while designing new systems. An extent of their influence depends on the experience of the engineering team, business requirements, available hosting infrastructure, and other project-specific factors.

\section{Related work}\label{section:related_work}

Given the popularity of SOA, many attempts have been made to re-think this software design approach and enhance it with better data handling capabilities. Götz et al. \cite{gotz2018challenges} proposes a data-driven approach towards designing microservices, which requires a complete data picture of a business as a prerequisite - something that can be difficult to acquire for any medium-to-large-sized business because of a sheer number of data flows and no built-in solutions to track them. Uber Engineering's Domain-Oriented Microservice Architecture \cite{gluck2020doma} can be seen as a step forward in this line of thinking, as this design approach only requires high-level understanding of main entities and the domains a business operates with. On a different note, Safina et al. extends Jolie, a programming language based on the microservices paradigm, in a way that allows to build services with Jolie in data-driven fashion \cite{safina2016data}. Z. Dehghani from technology consultancy company Thoughtworks proposed data meshes - a distributed data storage architecture that replaces monolithic data lakes \cite{dehghani2019move}, which improves data ownership decoupes data processing pipelines, but still hides data behind services. Contrary to this strand of software engineering research, rather than incrementally improving SOA, we wanted to consider using radically different approaches for better data management in software development.

The machine learning community is well aware of the concept of data flow. It is often used for implementation of ML tools. Notable examples include Tensorflow \cite{abadi2016tensorflow}, a general purpose library with focus on neural networks, Neuflow \cite{farabet2011neuflow}, a hardware architecture optimized for the computation of general-purpose vision algorithms, RLFlow \cite{liang2020distributed}, a reinforcement learning library, and more. The key difference between these tools and the analysis presented in this paper is that while the aforementioned tools are used to implement new ML algorithms, we explore general receptivity of FBP-driven software application to deployment of ML systems. Metaflow developed by Netflix \cite{netflix2020metaflow} shares our motivation described in Section~\ref{section:motivation} and implements many concepts discussed in the paper, and yet is also positioned solely as a tool for data science projects. We argue that there is certain benefit for considering Metaflow and similar tools in a broader scope of any data-processing applications.

A duality of control flow and data flow for building software was explored by the computer science community before \cite{treleaven1982towards, lauer1979duality, hasselbring2021control}. As a form of dataflow, FBP was contrasted with other paradigms and design principles, although never in the context of ML. In his book ``Flow-Based Programming: A new approach to application development'' the creator of FBP J.P. Morrison compares FBP and Object-Oriented Programming (OOP) \cite{morrison2010flow}. Similarly, Roosta contrasted dataflow and functional programming paradigms in his book ``Parallel Processing and Parallel Algorithms: theory and computation'' \cite{roosta2012parallel}. On a more practical side, Mironov et al.~\cite{mironov2021comparison} compared OOP and data-oriented approaches in the context of back-testing for trading strategies. They found the data-oriented approach is more efficient because of parallelism mechanisms. Lobunets and Krylovskiy implemented the same IoT processing system with SOA and FBP, and reported their experiences in regard to code coupling, debugging and testing \cite{lobunets2014applying}. While our methodology is similar to this work, our evaluation focuses on data and ML.
\section{Conclusions and future work}\label{section:conclusions}
This paper presents the results of an evaluation of FBP for ML deployment by comparing it against widely used SOA paradigm. We implemented four data-driven applications in both paradigms, completed a data science task for each of them, and measured evolution of the codebase throughout the process.

FBP is considered primarily a paradigm for distributed computing. Our experiment shows that FBP is a viable paradigm for developing general-purpose software according to DOA principles. We showed how FBP features (e.g., dataflow graph and data coupling) make data discovery and collection easier than in a traditional SOA setting. In addition, we highlighted how better tooling, that allows developers to define and traverse dataflow graphs at a higher level of abstraction, could help improve FBP adoption.

There are multiple ways to extend the experiment described in this paper. The same ML deployment perspective can be considered in the distributed context, where data streaming platforms (e.g., Apache Kafka) can be used. Observing evolution of a production DOA system implemented with FBP over a long period of time can quantify maintainability effects. Other paradigms (e.g., Actor model \cite{hewitt2010actor}) can be considered for comparison. Given that a lot of systems are already implemented with SOA, an analysis of full or partial migration from SOA to FBP for an existing system can be relevant.

We hope that our results highlight potential for FBP to become a standard paradigm for implementation of DOA systems. Exploration of approaches that promote better treatment of data is crucial to address challenges in ML deployment, considering the growing demand to leverage data for scientific and business purposes.

\bibliographystyle{ACM-Reference-Format}
\bibliography{references}
\end{document}